%% file: DH_model_generation.tex
\documentclass[review]{elsarticle}

\usepackage{lineno,hyperref}
\modulolinenumbers[5]

\usepackage{url}
\usepackage{comment}
\usepackage{todonotes}
\usepackage{amsmath}
\usepackage{nomencl}
\makenomenclature
\usepackage{etoolbox}
\usepackage{ulem}
\renewcommand\nomgroup[1]{%
	\item[\bfseries
	\ifstrequal{#1}{A}{Abbreviations}{      
		\ifstrequal{#1}{S}{Symbols}{ 			
			\ifstrequal{#1}{W}{Subscripts}{ 	    
				\ifstrequal{#1}{G}{Greek symbols}{}}}}  
	]}
\usepackage{float}
\definecolor{fzjgreen}{RGB}{185,210,95}
\definecolor{fzjyellow}{RGB}{250,235,90}
\definecolor{fzjred}{RGB}{235,95,115}
\definecolor{fzjlightblue}{RGB}{173,189,227}
\definecolor{fzjblue}{RGB}{2,61,107}
\definecolor{fzjgray}{RGB}{235,235,235}
\definecolor{fzjviolet}{RGB}{175, 130, 185}
\definecolor{fzjorange}{RGB}{250, 180, 90}
\usetikzlibrary{arrows.meta,arrows}

\usepackage{multirow}	
\usepackage{multicol}	
\usepackage{booktabs}	
\usepackage{dcolumn} 	
\newcolumntype{d}[1]{D{,}{,}{#1}}	
\newcolumntype{C}[1]{>{\centering\let\newline\\\arraybackslash\hspace{0pt}}m{#1}}
\usepackage{subcaption}
\usepackage{siunitx}
\DeclareSIUnit\annum{a}
\DeclareSIUnit\EUR{EUR}
\DeclareSIUnit\Wh{Wh}

\journal{arXiv}

\usepackage[capitalise, noabbrev]{cleveref}  

\begin{document}

\begin{frontmatter}

\title{Generation of Large District Heating System Models Using Open-Source Data and Tools: An Exemplary Workflow}


\author[a]{Jan Stock\corref{cor1}}
\ead{j.stock@fz-juelich.de}
\author[a]{Till Schmidt}
\author[a]{André Xhonneux}
\author[a,b]{Dirk Müller}

\cortext[cor1]{Corresponding author}
\address[a]{Institute of Climate and Energy Systems - Energy Systems Engineering (ICE-1), Forschungszentrum Jülich GmbH, Wilhelm-Johnen-Straße, 52425 Jülich}
\address[b]{Institute for Energy Efficient Buildings and Indoor Climate, E.ON Energy Research Center, RWTH Aachen University, Mathieustraße 10, 52074 Aachen}



\end{frontmatter}


\input{chapters/Introduction}

\input{chapters/ModelConstr_workflow}

\input{chapters/ModelConstr_example}

\input{chapters/Case_Studies}
\input{chapters/Discussion_Conclusion}

\section*{Author Contributions}
\textbf{Jan~Stock}: Conceptualization, Methodology, Writing - Original Draft, Writing - Review and Editing, Software, Visualization.
\textbf{Till~Schmidt}: Conceptualization, Methodology, Software, Visualization.
\textbf{André~Xhonneux}: Supervision, Writing - Review and Editing, Project administration, Funding acquisition, Resources.
\textbf{Dirk~Müller}: Supervision, Project administration, Funding acquisition, Resources. 

\section*{Declaration of competing interest}
The authors declare that they have no known competing financial interests or personal relationships that could have appeared to influence the work reported in this paper.

\section*{Acknowledgements}
This work was supported by the Helmholtz Association under the project “Helmholtz platform for the design of robust energy systems and their supply chains” (RESUR).


\input{DH_model_generation.bbl}
\appendix
\input{chapters/Appendix}

\end{document}

%% file: chapters/Introduction.tex
\section{Introduction}
\label{sec_introduction} 

District heating (DH) systems are a key technology for decarbonising the heat supply in the building sector. While new concepts such as low-exergy DH and cooling systems are being developed and established in newly built housing areas, existing DH systems must be transformed into a sustainable heat supply since most systems still rely on fossil-based heating plants. It is therefore essential to adapt DH systems currently in operation to achieve a sustainable heat supply infrastructure for the building sector. 

Existing DH systems need to be comprehensively analysed to identify the potential for integrating sustainable heat sources, e.g. renewable energy or waste heat sources, and to improve the DH operation efficiency. Efficiency improvements for existing DH systems can be achieved, for example, by reducing the DH supply temperature or using advanced control algorithms, which enable overall cost reductions and savings in CO\textsubscript{2} emissions~\cite{Li.2018, Lund.2018}. When analysing DH system adaptions, it is also becoming increasingly important to consider the power sector, as the coupling through power-to-heat technologies creates flexible multi-energy systems that enable the increasing use of fluctuating renewable energy sources~\cite{Bloess.2018}. However, to achieve the desired improvements, several measures may be necessary on the existing DH system infrastructure, such as replacing pipes due to changed mass flow conditions in the network or adapting buildings so that their internal heating systems can cope with lowered DH supply temperatures~\cite{Merlet.2023, Brange.2017, Guelpa.2023}. 

To analyse possible improvements in existing DH systems and identify possible required measures, data about an existing DH system is crucial. This data can be used in combination with various tools, such as simulation or optimisation models, for analytical purposes that enable the investigation of optimal design adaptions~\cite{Merlet.2023, Vannahme.2024} or the evaluation of advanced control approaches~\cite{Qin.2023, Hering.2022}. Furthermore, data on existing DH systems can be used to test developed analytical tools. For example, a developed DH network simulation tool can be evaluated on different DH systems with different network sizes and structures to test its applicability and to carry out benchmarks.

Although existing DH systems are categorised and localised in many cases~\cite{Pelda.2021, Muncan.2024, Triebs.2021, Manz.2024}, detailed data on individual DH systems, such as information on the connected buildings or DH operational data, is usually not publicly available.
Nevertheless, some data on the general structure of the pipe network is often available, e.g. DH operators offer potential DH customers the opportunity to check the supply by DH~\cite{IqonyGmbH.2024}. Moreover, general information about buildings is usually publicly available, e.g. in OpenStreetMap (OSM)~\cite{OpenStreetMapcontributors.2024}. These public data sources can be combined with software tools, which can estimate missing required information about DH systems through calculations, and thus enabling the generation of a digital representation of an existing DH system. Such a digital representation of a DH system, or called a DH model, can be used to analyse specific measures at the DH system, e.g. the utilisation of a sustainable heat source potential or the reduction of the DH supply temperature, or to test and evaluate developed analytical software tools for DH systems.

In this work, a workflow for generating a DH model using open-source data and tools is presented that can be used for general DH analyses and software tool testing. The general idea of the DH model generation workflow is introduced in \cref{sec_methodology_dh_model_construction} and a detailed example workflow is presented in \cref{sec_methodology_dh_model_construction_example}. Two generated DH models are shown in \cref{sec_case_studies}. Finally, \cref{sec_discussion_conclusion} discusses the workflow and the resulting DH models and summarises this work.

%% file: chapters/ModelConstr_workflow.tex
\section{Generation of district heating models}
\label{sec_methodology_dh_model_construction}

This section provides an overview of the workflow for generating a DH model using open-source data and combining it with open-source software tools to obtain a digital representation of a DH system. The management of the DH network data and the representation as a graph structure is done with the energy network management tool \textit{uesgraphs}~\cite{Fuchs.2016}. The workflow is demonstrated in an example use case in \cref{sec_methodology_dh_model_construction_example}.

\subsection{Workflow of generating district heating models}
\label{sec_methodology_dh_model_construction_overview}

A DH model refers to the digital representation of a DH system that includes information about the pipe network, heat sources and connected buildings. Such a DH model can therefore be used for analysing purposes, e.g. to create DH simulation models. 
As detailed comprehensive data of existing DH systems is usually not publicly available, the isolated data that is freely accessible from various sources is combined to obtain information about the most important properties of the DH system and to generate the best possible DH model. Depending on the DH system under investigation, the freely accessible data is quite heterogeneous in terms of the scope of information or the form of provision, e.g. the DH network structure can be provided as a simplified image-based map or as a file with a detailed geo-referenced graph structure. Missing information required for the DH model can be estimated using various calculation tools, e.g. the heat demand of the buildings supplied can be simulated using building simulation models. 

The basis of the DH model is a graph representation of the DH system with nodes for connected consumers and network forks, and edges for connecting pipes. In this graph structure, all information about the DH system is assigned to the corresponding graph elements, i.e. information about buildings is assigned to the corresponding building nodes, while pipe diameters or insulation classes are assigned to the corresponding edges in the graph. Such a graph representation is often used in DH system analyses, e.g. for simulation studies~\cite{Fuchs.2016, Nord.2021, Gross.2021, Han.2023} 
or topology analyses~\cite{Zhong.2020, Schmidt.2021, Salenbien.2023, Li.2022}. 

\begin{figure}[htb!]
	\centering
	\includegraphics[width=\linewidth]{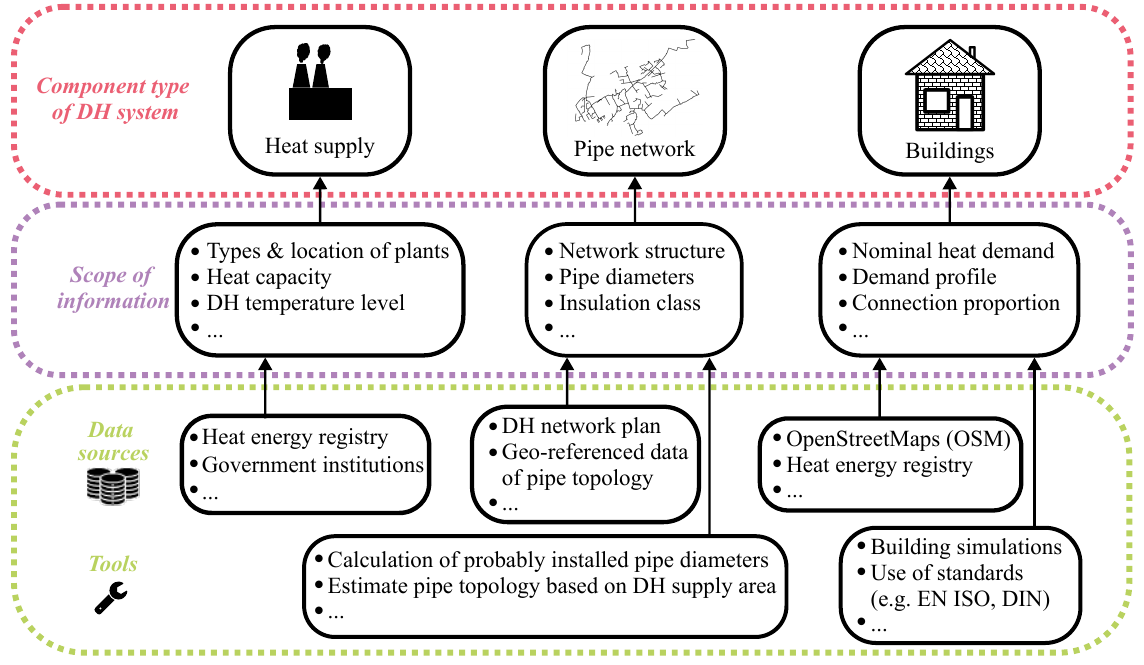}
	\caption{Overview of generating a comprehensive DH model. The three main types of DH components are assigned the required information, which is aggregated from data sources or calculated by tools.}
    \label{fig_model_generation}
\end{figure}

The general workflow for generating a graph-based DH model from several open-source data sets, supplemented by open-source tools, is presented in \cref{fig_model_generation}. The three main DH component types of the DH model, the heat supply, the pipe network and the supplied buildings (first, red level), are assigned the corresponding information (second, purple level), which either comes from data sources or is estimated using calculation tools (third, green level).

\subsection{Define the aim of application and required information}
\label{}
First, the objective of using the DH model must be specified to define the required scope of information (see second level of \cref{fig_model_generation}), i.e. what information about the DH system is needed. The third, green level of \cref{fig_model_generation} describes some exemplary data sources for the three types of DH components. For example, information about the network topology is sufficient for a general network analysis, e.g. spatial analysis of the heat source potential concerning the existing DH network or the identification of possible DH network extension areas. However, detailed network simulations require additional information regarding the pipe diameters and insulation classes to calculate the hydraulic and thermal conditions in the network. To simulate the DH system, the buildings connected to the DH network and their heat demand profiles must also be known, as exemplarily listed for the building component in \cref{fig_model_generation}. However, suppose the possibilities of replacing or decarbonising the heat supply are analysed, information about the currently operated heating plants in terms of location, plant type and capacity is usually sufficient. 

\subsection{Examples for data sources and tools}
\label{}
The third level of \cref{fig_model_generation} shows some examples of open-source data sets or tools from which information about the DH system can be collected or estimated. For example, heat energy registries are promising data sources as they provide data for local heating planning by including information on existing heating plants, heat source potential or information on building characteristics and heat demand. 
If the required data about the DH system is not available for the intended DH model application, the exemplary tool approaches can be used to estimate this information. In general, data sources and tools can be substituted arbitrarily, which also depends on what data is available for the DH system under investigation. 

\begin{figure}[htb!]
	\centering
	\includegraphics[width=\linewidth]{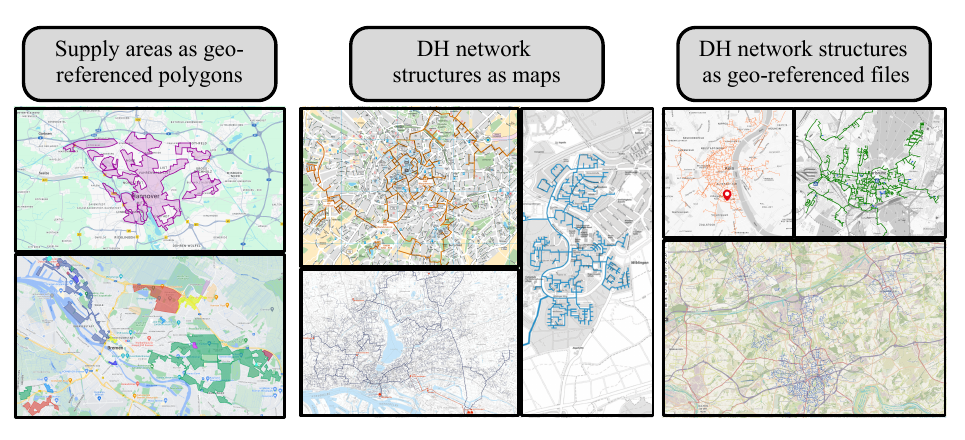}
	\caption{Examples of available information on DH network structures. Supply areas of DH systems: top Hannover~\cite{enercityAG.2024}, bottom Bremen~\cite{wesernetzBremenGmbH.2024}; maps of DH networks: top left Aachen~\cite{STAWAGStadtundStadteregionswerkeAachenAG.2011}, bottom left Hamburg~\cite{HausundGrundeigentumervereinHamburgRahlstedte.V..2023}, right Ulm (subnetwork)~\cite{FernwarmeUlmGmbH.2024}; visualised geo-referenced files of DH networks: top left Köln~\cite{RheinEnergieAG.2024}, top right Karlsruhe~\cite{StadtwerkeKarlsruheGmbH.2024}, bottom Bottrop, Gelsenkirchen and Essen~\cite{IqonyGmbH.2024}. 
 }
	\label{fig_network_structure_examples}
\end{figure}

For many DH system analyses, data about the existing DH network is crucial. DH supply areas or network topologies are often published by DH operators to offer potential new DH customers a service to check the possibility of DH supply for their building. The data is usually provided online in different ways, as shown in \cref{fig_network_structure_examples}. Sometimes only supply areas are given, while other DH operators provide detailed DH network structures as maps or even as geo-referenced network files that are embedded in the website of the DH operator.

All these available data sources, presented in \cref{fig_network_structure_examples}, can be used to obtain information about the existing DH network structure. If a supply area is made available, the network structure can be estimated, as the pipes are typically installed along the existing street network. Based on the given location of a heating plant, a pipe network can be designed along the street network, connecting the buildings along the streets according to a specified DH connection rate. For example, \textit{uesgraphs} provides such a method for designing a DH network structure in a specific region~\cite{Fuchs.2016}. 

If the DH network structure is available in an image-based map, the data can be extracted using pixels or vector graphic elements. The map is referenced with geo-information by defining several geo-reference points in the map. The data in the map is then transformed into the corresponding coordinate system depending on the assigned geo-reference points. Afterwards, the network structure on the map is identified by determining the colour code of the displayed network, e.g. the correct RGB blue colour code is determined when the network is displayed in blue on the map (e.g. see \cref{fig_network_structure_examples} DH network of Ulm). Since the pixels or vector graphic elements are mapped to geo-coordinates and the network information is identified by colour-coding, a graph-based network structure can be generated and used for DH model generation.

The identification of the DH network is most straightforward when kml-files are available
\footnote{Keyhole Markup Language (KML) is a markup language for describing geo data.}.
These files are embedded in the website of the DH operators and thus can be downloaded. Since the network structure in the kml-file is already geo-referenced, the graph-based DH network is directly accessible. However, a reformulation of the data structure for the desired graph representation tool, e.g. \textit{uesgraphs}, might be necessary.

If no detailed data is available, tools can be used to calculate missing information or estimate parameters. For example, design standards can be applied to estimate probable pipe diameters of the individual segments of the DH network (e.g. see \cref{sec_methodology_dh_model_construction_example}).

For the buildings supplied by the DH system, detailed heat consumption data, particularly resolved over time, is usually not publicly accessible due to data protection of the building occupants. Therefore, simulation models can serve as a replacing tool to calculate heat demand profiles of buildings (e.g. \cite{Remmen.2018}) and thus estimate the heat demand of the DH model. The data required to build a building simulation model can be obtained, for example, from OSM data which contains detailed geo-referenced building data and information about the building net leased area, the usage type or sometimes the year of construction, which helps to estimate the heat demand.

%% file: chapters/ModelConstr_example.tex
\section{Example of district heating model generation}
\label{sec_methodology_dh_model_construction_example}

The generation of a DH model is exemplarily shown by modelling the DH system of Bottrop. The city of Bottrop is located in the federal state of North Rhine-Westphalia (NRW), Germany. 
At first, the intended analysis objective for which the DH model is to be used must be defined to determine what data is required about the DH system. In this case, the DH model is used to analyse a possible DH network separation, in which an existing DH system is separated to partially transform the separated DH network part depending on the local conditions of the network and supplied buildings, taking into account a newly utilised heat source~\cite{Stock.2024, Stock.2024c}. In addition, the separated and remaining DH network are simulated to evaluate the feasible operation of both arising DH systems. Therefore, the following information about the Bottrop DH system is required to carry out DH network separation analysis: the DH network structure, the pipe diameters, information about the locations of the heating plants, and information about the buildings. The buildings actually supplied in Bottrop must be identified and their heat demand determined over time, whereby their year of construction is also important for estimating the required heating system temperatures. For more detailed information on the analysis of DH network separation using the generated DH model, it is referred to~\cite{Stock.2024c}.

\subsection{Pipe network}
\begin{figure}[htb!]
	\centering
	\includegraphics[width=\linewidth]{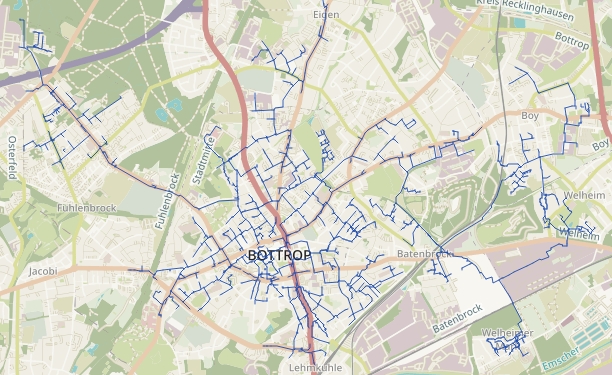}
	\caption{Network structure of the Bottrop DH system~\cite{IqonyGmbH.2024}.}
	\label{fig_network_structure_bottrop}
\end{figure}

The general topology of the Bottrop DH network shown in \cref{fig_network_structure_bottrop} is publicly accessible as a kml-file embedded in the website of the DH operator~\cite{IqonyGmbH.2024} and is imported as a graph object in \textit{uesgraphs}. The kml-file contains information about the general pipe structure of the DH network, but it does not include details about the connected buildings, heating plants or pipe diameters.

\subsection{Buildings}
\label{sec_methodology_dh_model_construction_example_bldg}
The information on the buildings supplied by the DH system comes from the heat energy registry "Energieatlas NRW" of the federal state NRW~\cite{LandesamtfurNaturUmweltundVerbraucherschutzNRW.2024}, which provides comprehensive building information. The Energieatlas NRW data set contains a heat demand model that estimates the annual heat demand of the geo-referenced buildings in NRW based on their known footprint, their usage type and local weather conditions~\cite{LandesamtfurNaturUmweltundVerbraucherschutzNRW.2024}. Based on this annual heat demand, the Python tool \textit{demandlib}~\cite{oemofDeveloperGroup.2016} is used to create time-based heat demand profiles for each building, considering standard load profiles for different building types. 

\begin{figure}[htb!]
	\centering
	\includegraphics[width=\linewidth]{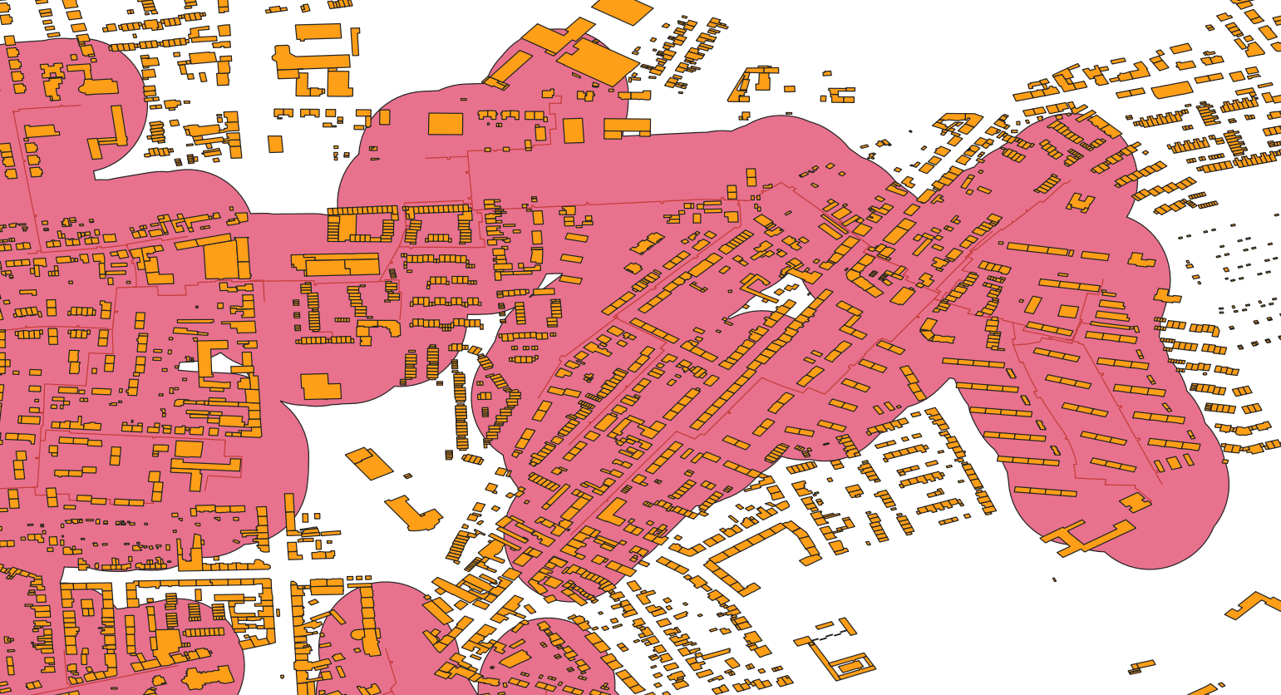}
	\caption{Example of a buffered network structure. The polygons marked orange represent the existing buildings in this region.}
	\label{fig_network_buffered}
\end{figure}

It is important to consider that not every building in a city is connected to the local DH system. 
First, only buildings in close proximity to the DH network structure are considered. Therefore, the graph object of the DH network is buffered to take into account the buildings in the vicinity of the network with a certain threshold. The threshold, i.e. the length of the vector forming the buffer around the network, is set to 100~m. With this threshold, the buildings in Bottrop are sufficiently filtered for the next step, in which the DH connection proportion is taken into account. The buffering of the network structure is exemplarily illustrated in \cref{fig_network_buffered}. The buildings that are not within the buffered network area are not further considered. 

\begin{figure}[htb!]
	\centering
	\includegraphics[width=\linewidth]{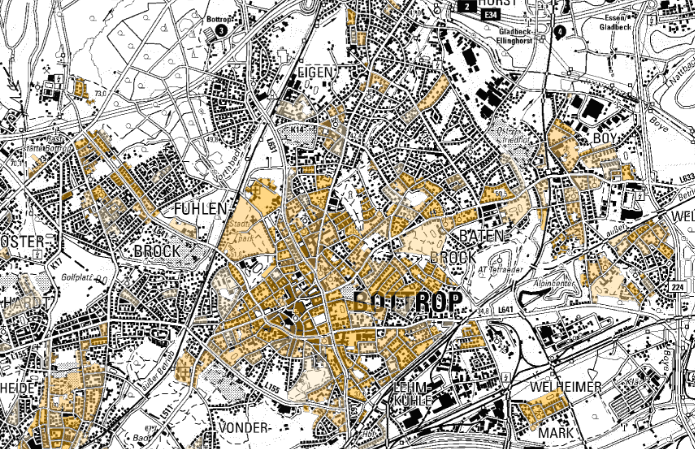}
	\caption{DH connection proportion in the city of Bottrop~\cite{LandesamtfurNaturUmweltundVerbraucherschutzNRW.2024}.}
	\label{fig_dh_connection_proportion}
\end{figure}

Second, the buildings actually supplied in the vicinity of the DH system must be fixed. Information on the actual local DH connection proportion, i.e. the proportion of buildings supplied by DH in a specific area, is available in the Energieatlas NRW~\cite{LandesamtfurNaturUmweltundVerbraucherschutzNRW.2024} (see \cref{fig_dh_connection_proportion}) and provided by the project-executing institute in a detailed resolution~\cite{InWISInstitutfurWohnungswesenImmobilienwirtschaftStadtundRegionalentwicklungGmbH.2021}. The provided data set is resolved locally for building blocks. For each building block that is within the previously defined buffered network area, the number of buildings representing the connection proportion are randomly selected and connected to the DH system, i.e. added to the graph object. For building blocks for which no data on the DH connection proportion is available, all buildings in the respective block are connected to the DH network. 

An additional data source is used to assign a year of construction to the buildings supplied by DH, e.g. to estimate the required supply temperature of the building heating system. The year of construction of buildings is given in the German census~\cite{StatistischesBundesamt.2011}. In this open-source data set, the year of construction of several buildings is reported and statistically extrapolated to a 100x100~m grid. The year of construction of all buildings connected to the DH network is assigned according to the year of construction specified in the 100x100~m grids provided by~\cite{StatistischesBundesamt.2011}. In this exemplary case, 80~\% of the required 100x100~m grids are available. If no data is available for a particular building, the average year of construction of the neighbouring buildings is calculated and assigned.

\subsection{Heat supply}
\begin{figure}[htb!]
	\centering
	\includegraphics[scale=0.2]{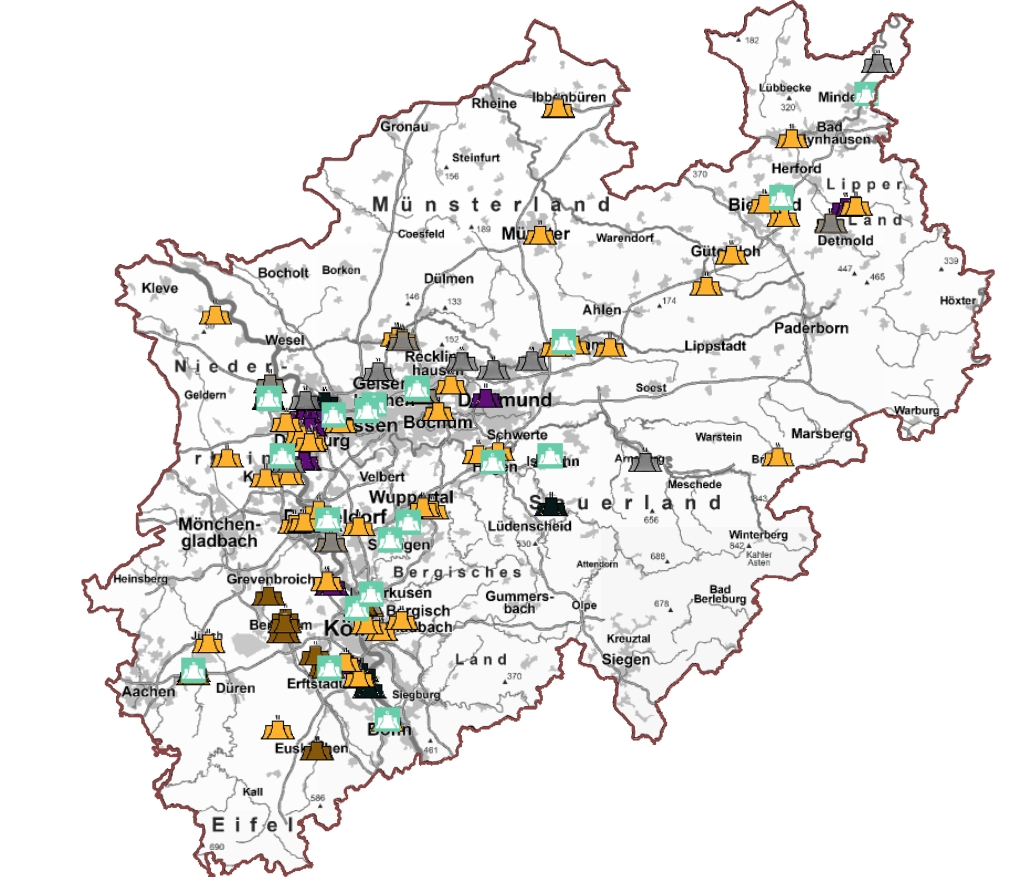}
	\caption{A selection of heating plants in NRW~\cite{LandesamtfurNaturUmweltundVerbraucherschutzNRW.2024}.}
	\label{fig_heating_plants}
\end{figure}

Several heating plants and co-generation plants in NRW are listed in the Energieatlas NRW (see \cref{fig_heating_plants}), but are not assigned to a specific DH system~\cite{LandesamtfurNaturUmweltundVerbraucherschutzNRW.2024}. Therefore, the list of all plants in the region of the modelled DH system is checked to connect all heating plants close to the DH network to the graph object as a supply node.

\subsection{Pipe diameters}
Since the pipe diameters of the DH network are not known, the diameters of the pipes are calculated by taking the required mass flows into consideration. Based on the nominal heat demand of each supplied building and an assumed temperature difference of 30~K at the building substations, the required nominal mass flow for each building can be calculated. As the pipe routing of the DH network and the location of the heating plant are known, the required diameters can be calculated using design standards~\cite{Nussbaumer.April2017}. This approach of determining DH pipe diameters is also included in \textit{uesgraphs}. In this way, the likely installed pipe diameters are calculated and assigned to the DH model.

\subsection{Model size reduction by clustering}
Finally, the generated DH model can be optionally simplified by spatial node aggregation, i.e. reducing the model size for complex model analyses such as dynamic simulation models~\cite{Guelpa.2019, Falay.2020} or optimisation models~\cite{Kotzur.2021}. Therefore, a clustering approach can be applied to the building areas to create aggregated consumer nodes in the graph object that represent multiple buildings. The k-means algorithm is chosen for this purpose, as the main network structure is not changed by this algorithm and a certain number of resulting clusters can be specified. The total number of buildings within a consumer node depends on the spatial conditions, i.e. how many buildings are in close proximity to each other. A consumer node aggregates the heat demand profiles of all buildings it contains, while the year of construction of the consumer node is the average year of construction of all buildings present. 

The various steps described for setting up the DH model, starting with a graph representation of the network, the assignment of building nodes to the graph and, optionally, clustering of the building nodes to aggregated consumer nodes, are visually summarised for the Bottrop DH system in \ref{sec_appendix_dh_model}.

%% file: chapters/Case_Studies.tex
\section{Generated district heating models}
\label{sec_case_studies}

\subsection{District heating system of Bottrop}
\begin{figure}[htb!]
	\centering
	\includegraphics[width=\linewidth]{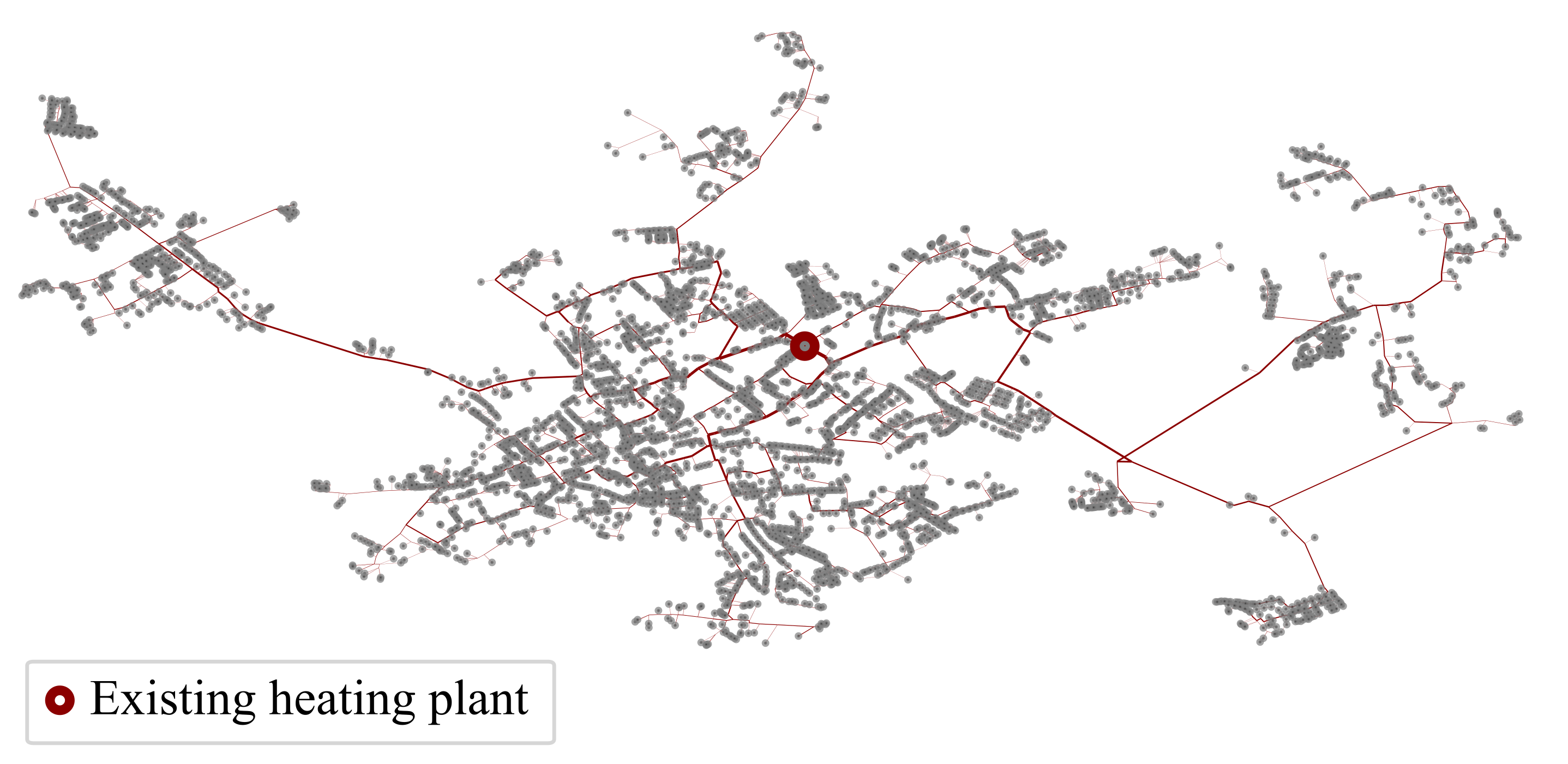}
	\caption{DH model of the Bottrop DH system with all identified buildings connected and designed pipe diameters.}
	\label{fig_bottrop_dh_network}
\end{figure}

The first DH model presented is the DH system of Bottrop, which is also used to illustrate the various steps of the DH model generation workflow in~\cref{sec_methodology_dh_model_construction_example}. 4,458 buildings are identified to be supplied by the DH network. The network structure is characterised by many tree-shaped elements, while one central heating plant supplies the DH system. The resulting DH network structure with all supplied buildings is shown in \cref{fig_bottrop_dh_network}. The width of the pipes symbolises the pipe diameter.

\subsection{District heating system of Essen}
\begin{figure}[htb!]
	\centering
	\includegraphics[width=\linewidth]{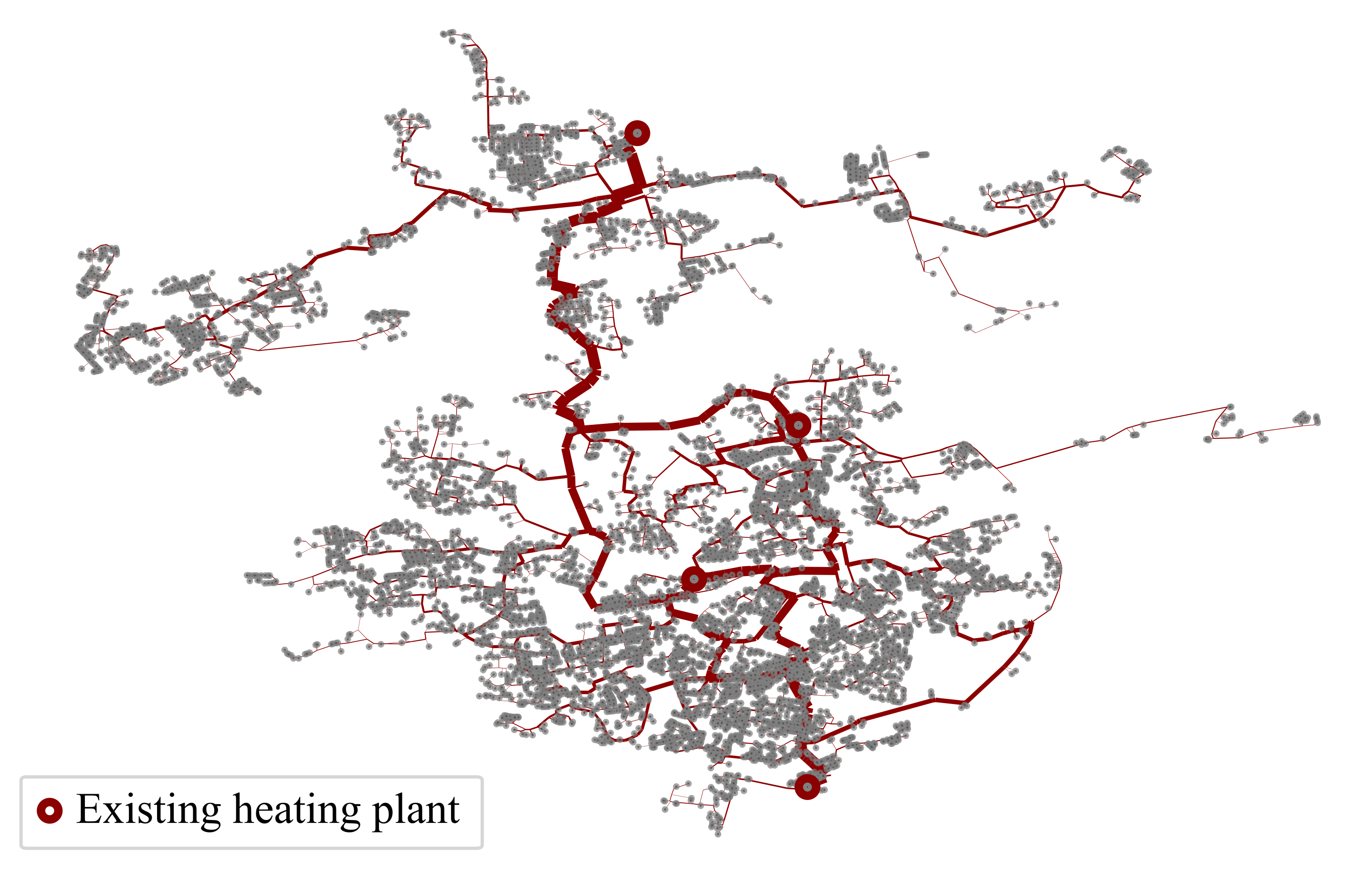}
	\caption{DH model of the Essen DH system with all identified buildings connected and designed pipe diameters.}
	\label{fig_essen_dh_network_8000}
\end{figure}

\begin{figure}[htb!]
	\centering
	\includegraphics[width=\linewidth]{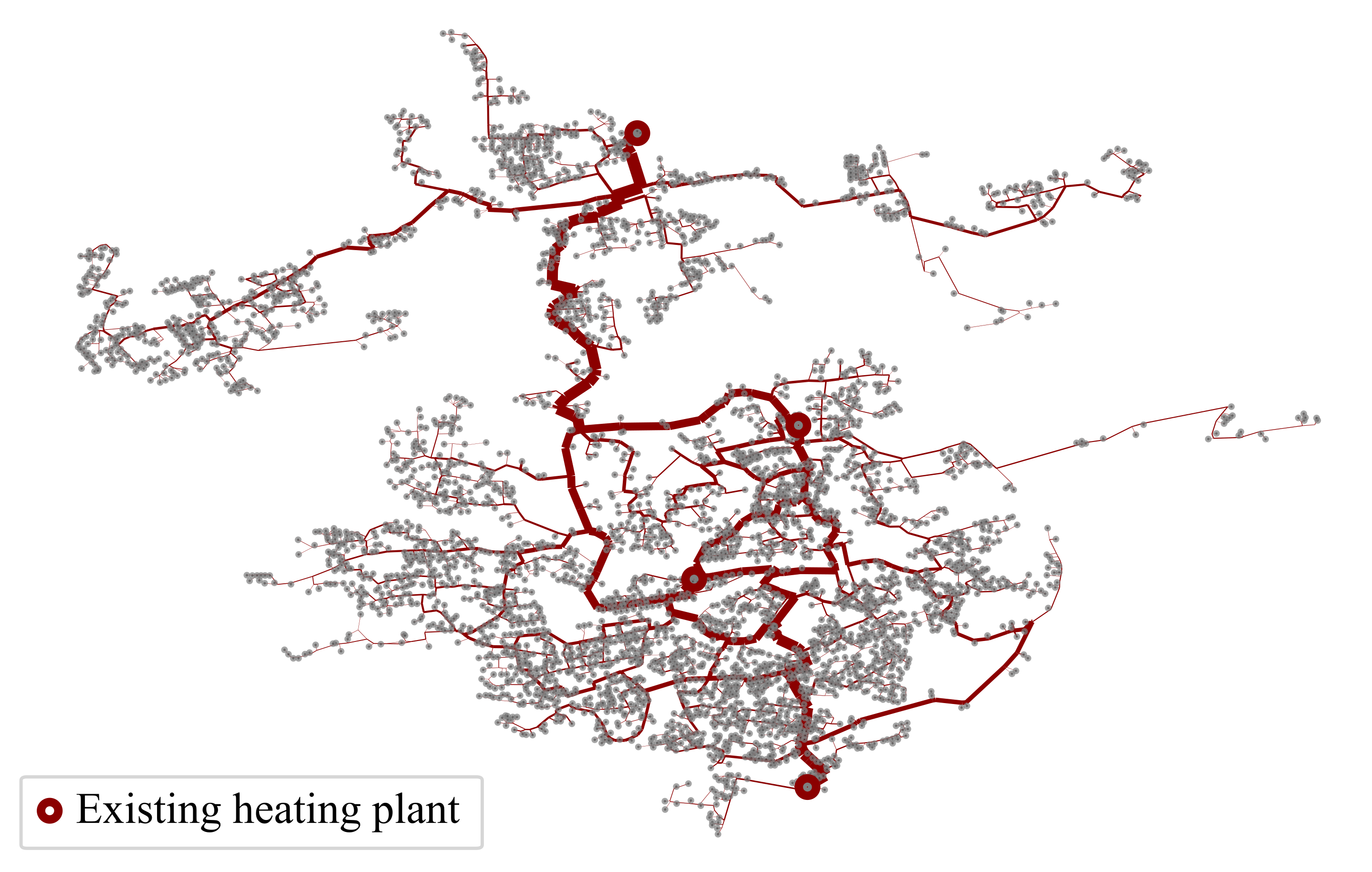}
	\caption{DH model of the Essen DH system with the buildings clustered to 4,000 consumer nodes and designed pipe diameters.}
	\label{fig_essen_dh_network_4000}
\end{figure}

The second example of a DH model is the DH system of Essen, where the building clustering approach is applied. Since the DH system of Essen is close to the Bottrop DH system, the available data and the generation workflow of the DH model are similar to those described in \cref{sec_methodology_dh_model_construction_example}.  The Essen DH system has a much more meshed network structure than the Bottrop DH network and four operating heating plants. As the DH system is much larger than the one in Bottrop, with 8,066 buildings identified in the DH model generation workflow, the building nodes are clustered into 4,000 aggregated consumer nodes in the network graph to reduce the size of the DH model and thus to enable the application of complex network simulations or optimisation models. The resulting DH network is shown below, first non-clustered with all 8,066 buildings connected in~\cref{fig_essen_dh_network_8000} and second with the building nodes being clustered to 4,000 consumer nodes in \cref{fig_essen_dh_network_4000}.

%% file: chapters/Discussion_Conclusion.tex
\section{Discussion and Conclusion}
\label{sec_discussion_conclusion} 

The workflow of DH model generation using open-source data and tools is an option for creating useful models of existing DH systems when no real data is available. The generated DH models can be used either to analyse DH measures and comprehensive transformation pathways for the corresponding DH system or for testing developed analysis tools. The development of this workflow is motivated by generating usable DH model data to test a developed tool for DH network separation in~\cite{Stock.2024c}, which also represents a possible transformation pathway as the analysed DH system is partially transformed to utilise a sustainable heat source. 
In general, the depth of information required depends heavily on the objective of DH model application, since only information about the network structure is needed for a general topology analysis, while detailed simulations require more precise data about the pipe characteristics and the connected buildings. 

The different quality of the available data and the assumptions made during the generation workflow influence the resulting DH model. For example, the DH network can be modelled more accurately if a geo-referenced kml-file is available than if the pipe network in a DH supply area is estimated. But even if detailed data is available, assumptions are necessary that can influence the DH model, e.g. the selection of buildings in a building block to achieve the given DH connection proportion in the corresponding building block (see \cref{sec_methodology_dh_model_construction_example_bldg}). 
Furthermore, some data sources may be outdated, e.g. the DH system may have been expanded or reconstructed since the data was recorded. The data sources used may also be incomplete, e.g. some buildings or heating plants currently in operation may not be included in the data sets used.
Although assumptions and missing data may influence the resulting DH model, the objective of applying the DH model must be considered, e.g. assessing the heat source potential in the surrounding of the DH system is not as sensitive to the DH model quality as analysing the replacement of individual pipe segments. Nevertheless, the assumptions made and their influence on the resulting DH model must be taken into account when using the DH model for analysis, i.e. plausibility checks and validations of the generated DH model with the real DH system are crucial to derive reliable results from DH model analysis.

The DH models generated in this work use various available information, from the DH operator and the local heat energy registry, but also tools to estimate the heat demand of supplied buildings or the diameters of the laid pipes. Both DH systems modelled represent large existing DH systems with thousands of supplied buildings. The largest modelled DH system in Essen, with over 8,000 connected buildings, is exemplarily clustered to reduce the size of the DH model and thus to enable the application of complex simulation and optimisation tools.
However, the actual quality of the created DH models compared to the real existing DH systems can not yet be validated because no real data is available. However, the intended use of these DH models in the current development stage is the tool testing for DH network separation~\cite{Stock.2024c}. In this context, the generated DH models are usable because the most important information about network structure and the local buildings supplied is available, which have the greatest influence on the decisions of network separation. Overall, the created DH models do not claim to be correct, but they should reflect the dimensions of large DH systems that supply several thousand buildings, which the resulting DH models achieve. 
    
Nevertheless, future work should focus on validating the generated DH models by working with DH operators who can provide real data about the DH system. Furthermore, the presented workflow is being expanded to increase the variety of open-source data sets that can be used to generate DH models.

%% file: chapters/Appendix.tex
\clearpage
\section{Visually representation of the Bottrop DH system for different DH model construction steps}
\label{sec_appendix_dh_model}

\begin{figure}[htb!]
    \centering
    \includegraphics[width=\linewidth]{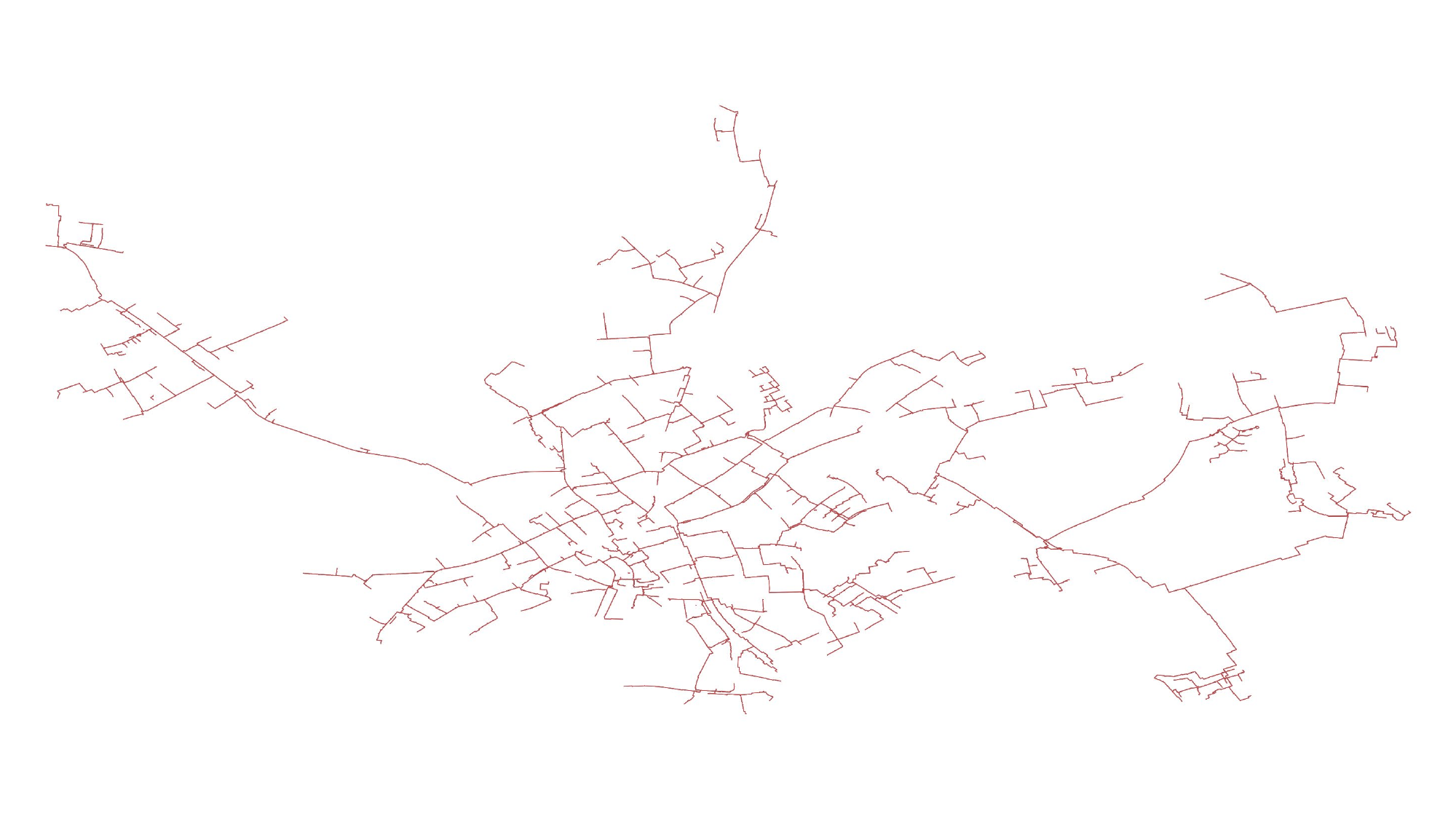}
    \caption{General DH network structure.}
    \label{fig_dh_model_step_1}
\end{figure}

\begin{figure}[htb!]
    \centering
    \includegraphics[width=\linewidth]{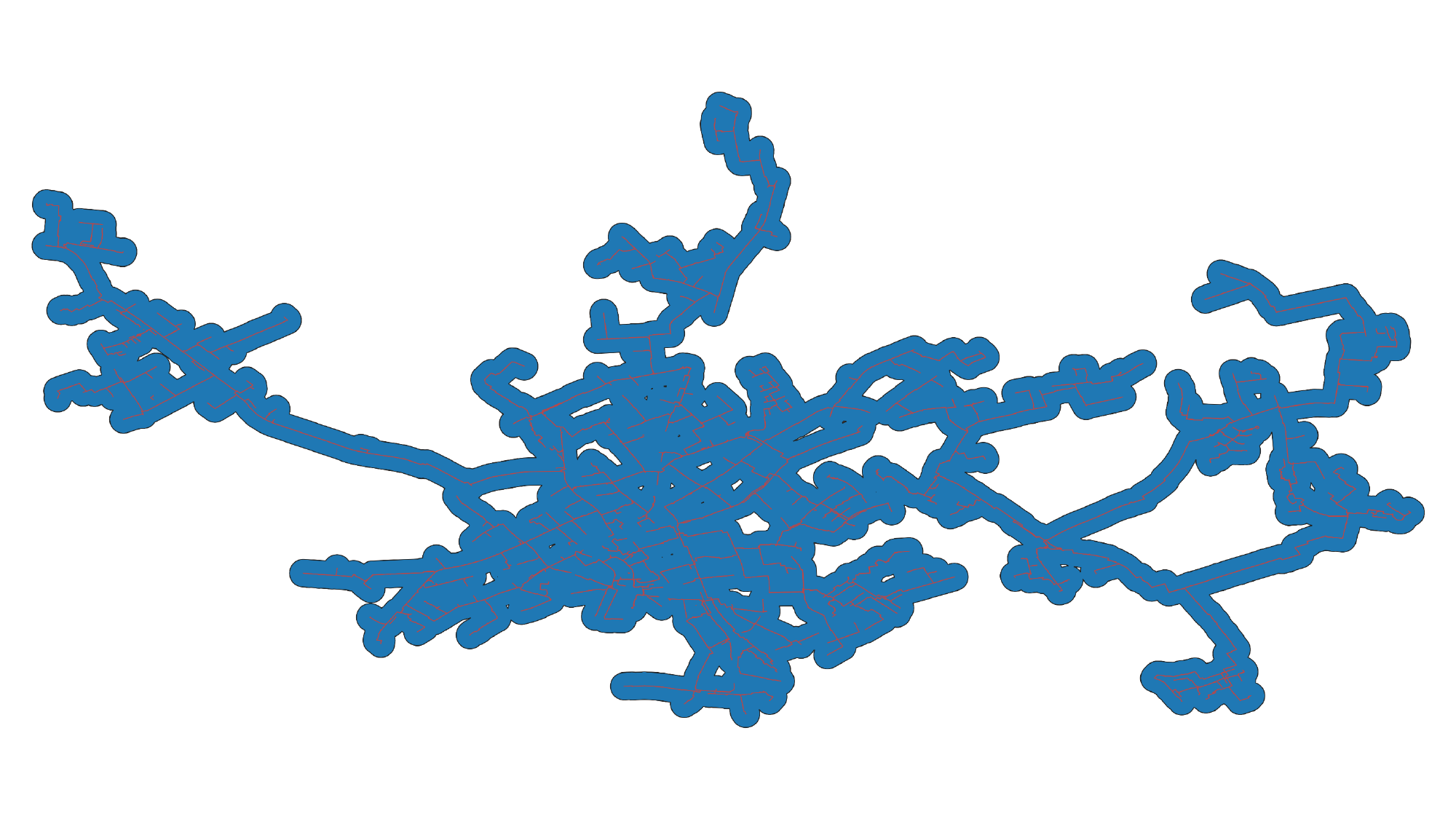}
    \caption{Buffered DH network structure.}
    \label{fig_dh_model_step_2}
\end{figure}


\begin{figure}[htb!]
    \centering
    \includegraphics[width=\linewidth]{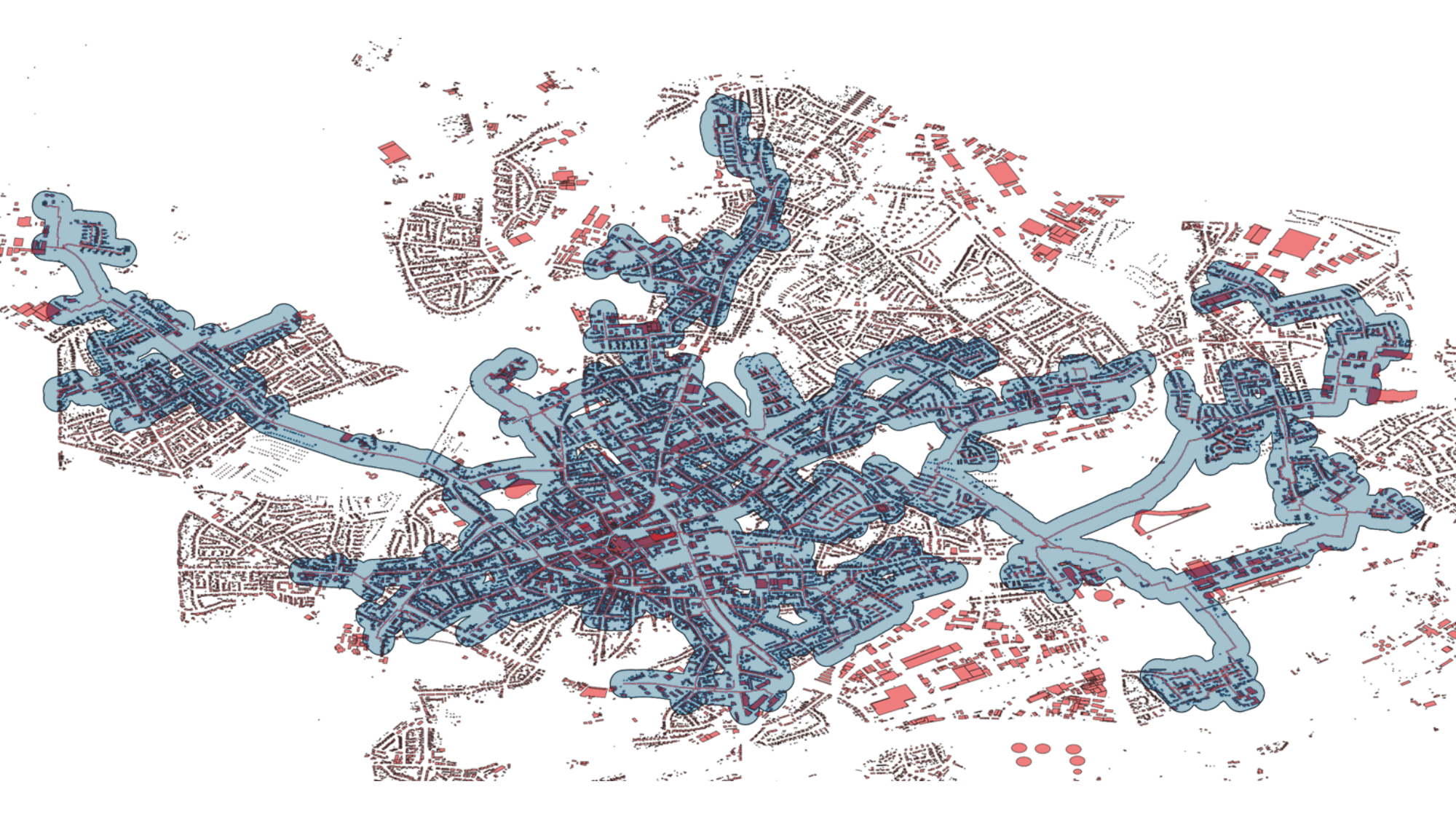}
    \caption{Combined buffered DH network structure and geo-referenced building locations.}
    \label{fig_dh_model_step_4}
\end{figure}

\begin{figure}[htb!]
    \centering
    \includegraphics[width=\linewidth]{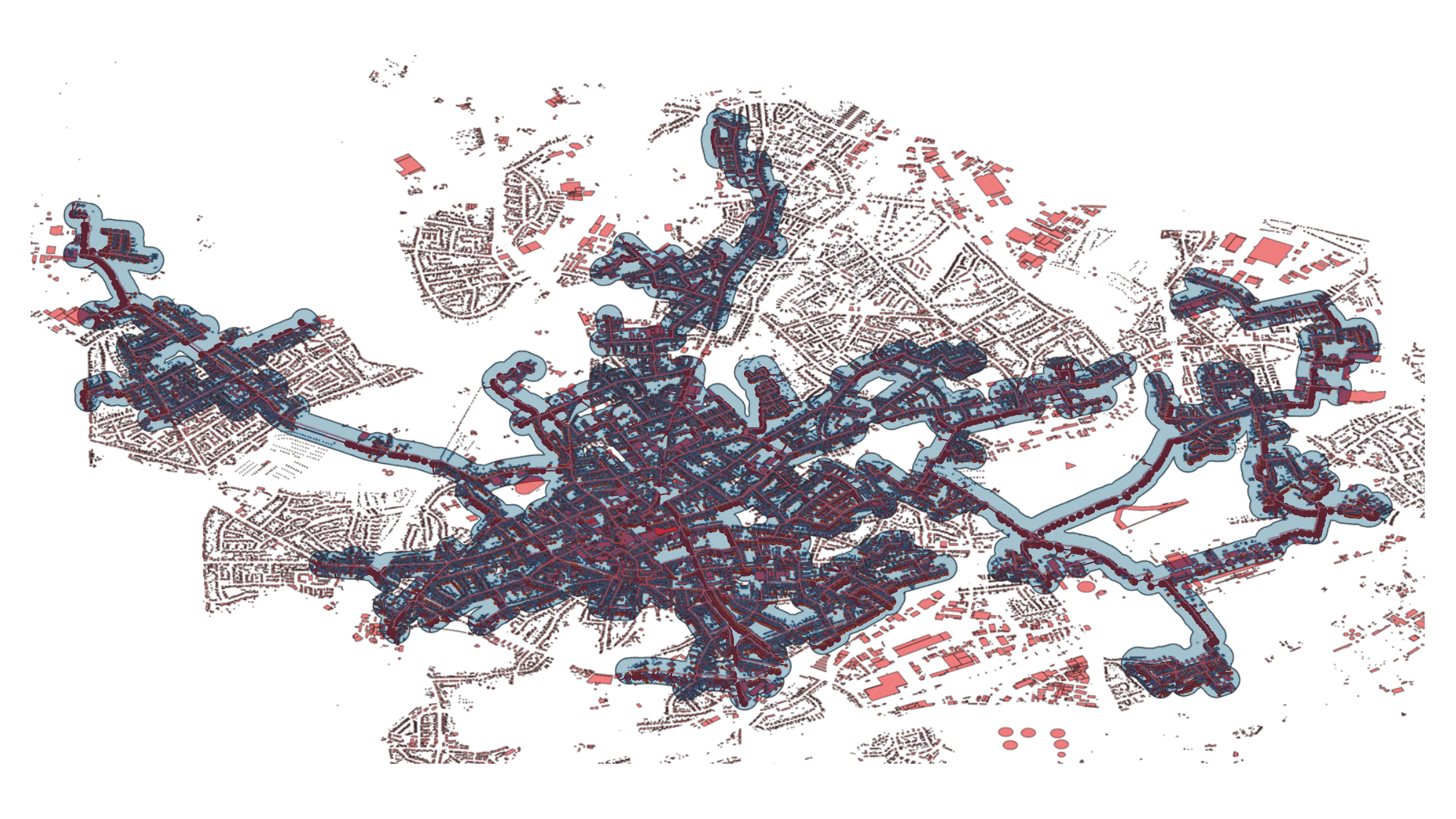}
    \caption{Buffered DH network structure and connected buildings to the DH network.}
    \label{fig_dh_model_step_5}
\end{figure}

\begin{figure}[htb!]
    \centering
    \includegraphics[width=\linewidth]{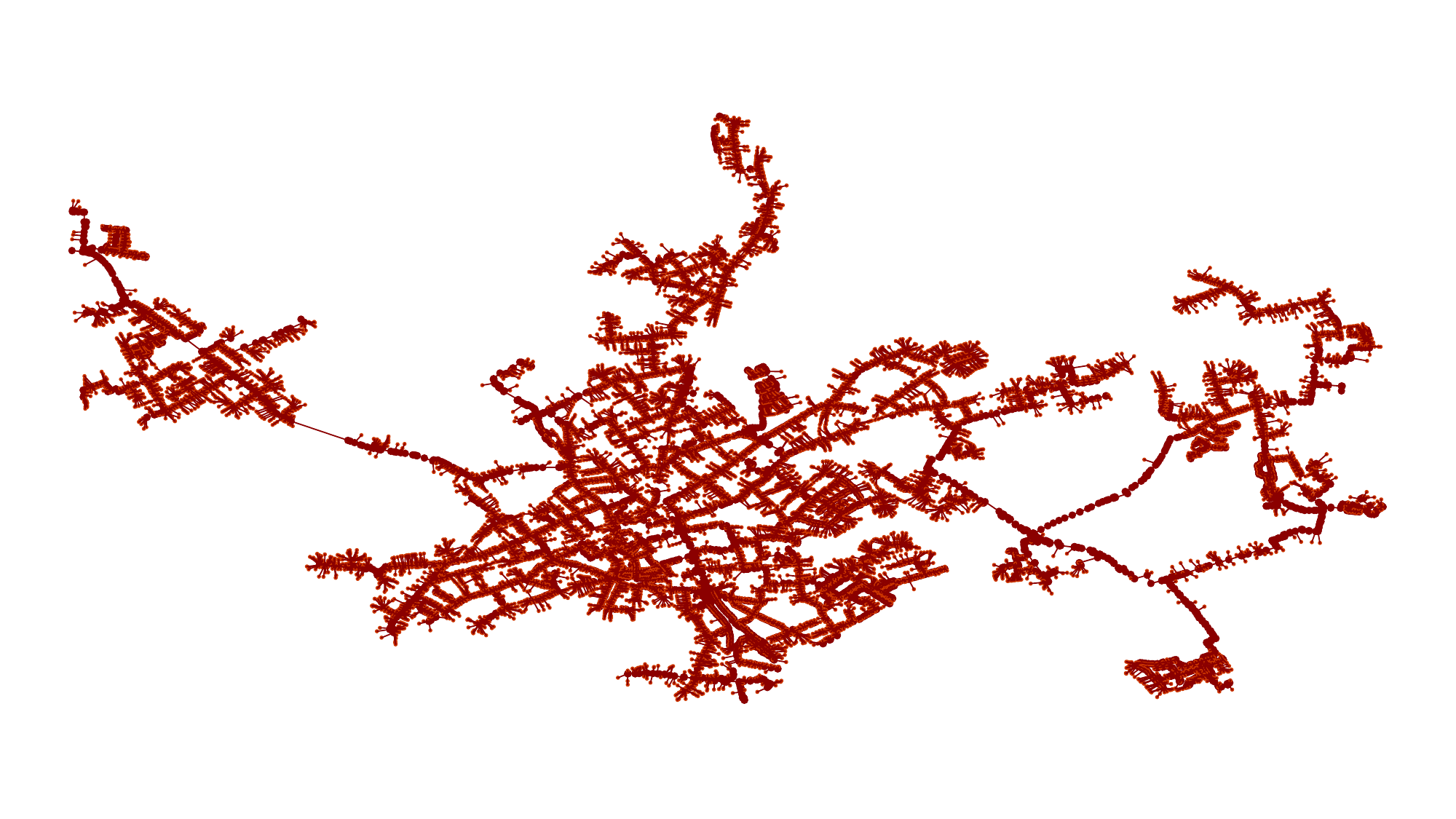}
    \caption{DH network structure with connected buildings.}
    \label{fig_dh_model_step_6}
\end{figure}

\begin{figure}[htb!]
    \centering
    \includegraphics[width=\linewidth]{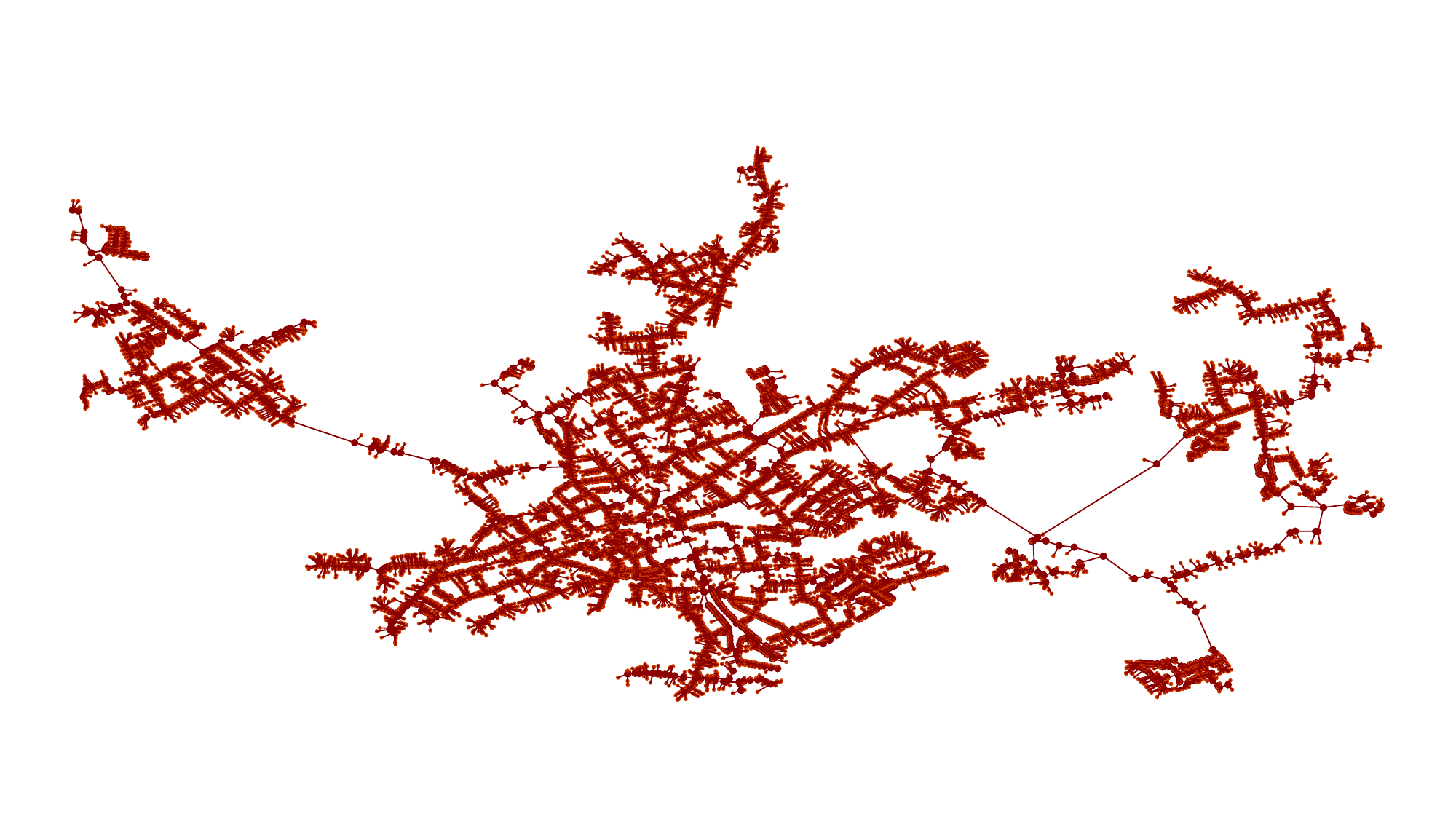}
    \caption{Simplified DH network structure with connected buildings; the network nodes that only connect two edges are removed.}
    \label{fig_dh_model_step_7}
\end{figure}

\begin{figure}[htb!]
    \centering
    \includegraphics[width=\linewidth]{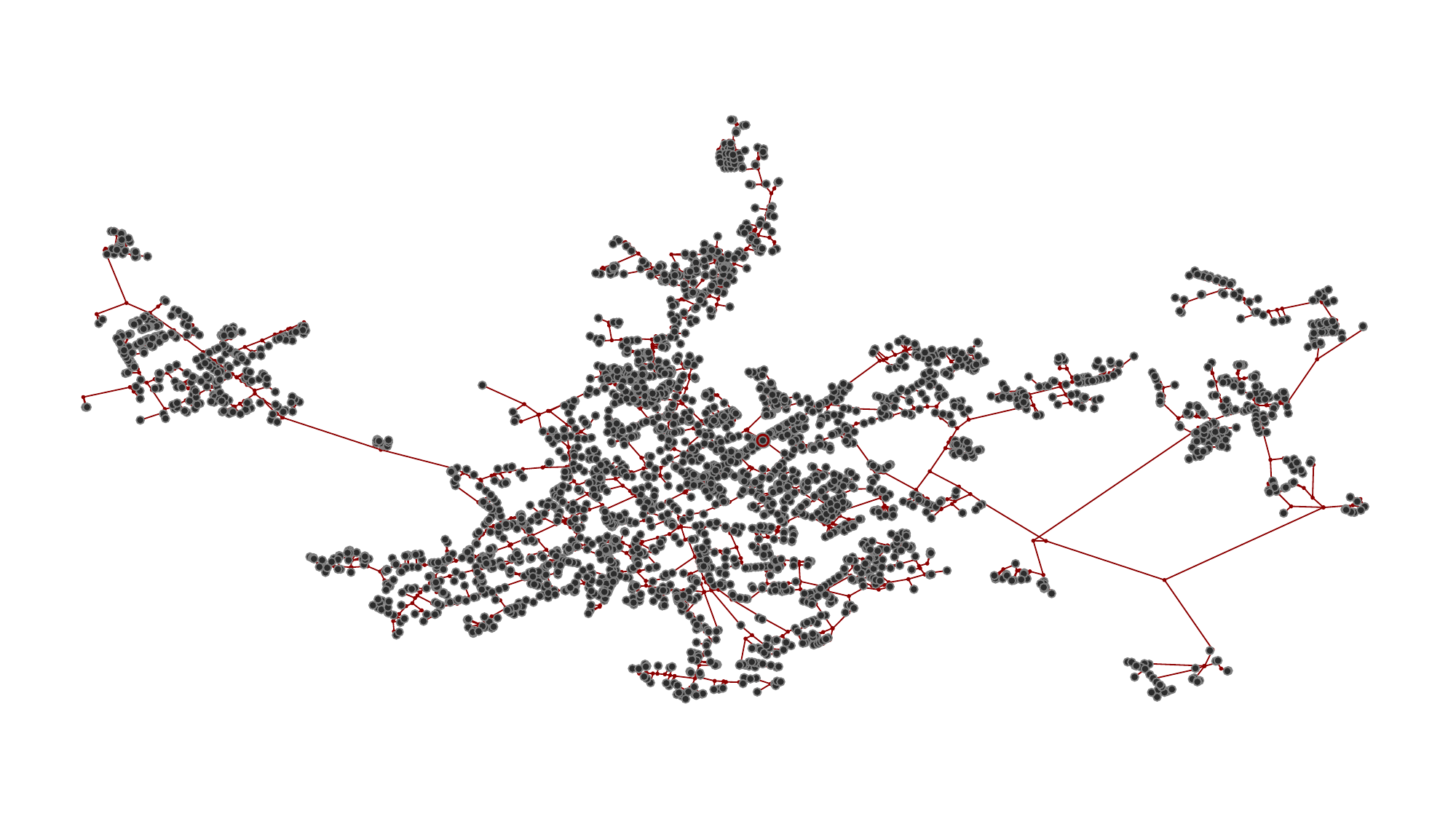}
    \caption{Clustering building nodes to aggregated consumer nodes.}
    \label{fig_dh_model_step_8}
\end{figure}

\begin{figure}[htb!]
    \centering
    \includegraphics[width=\linewidth]{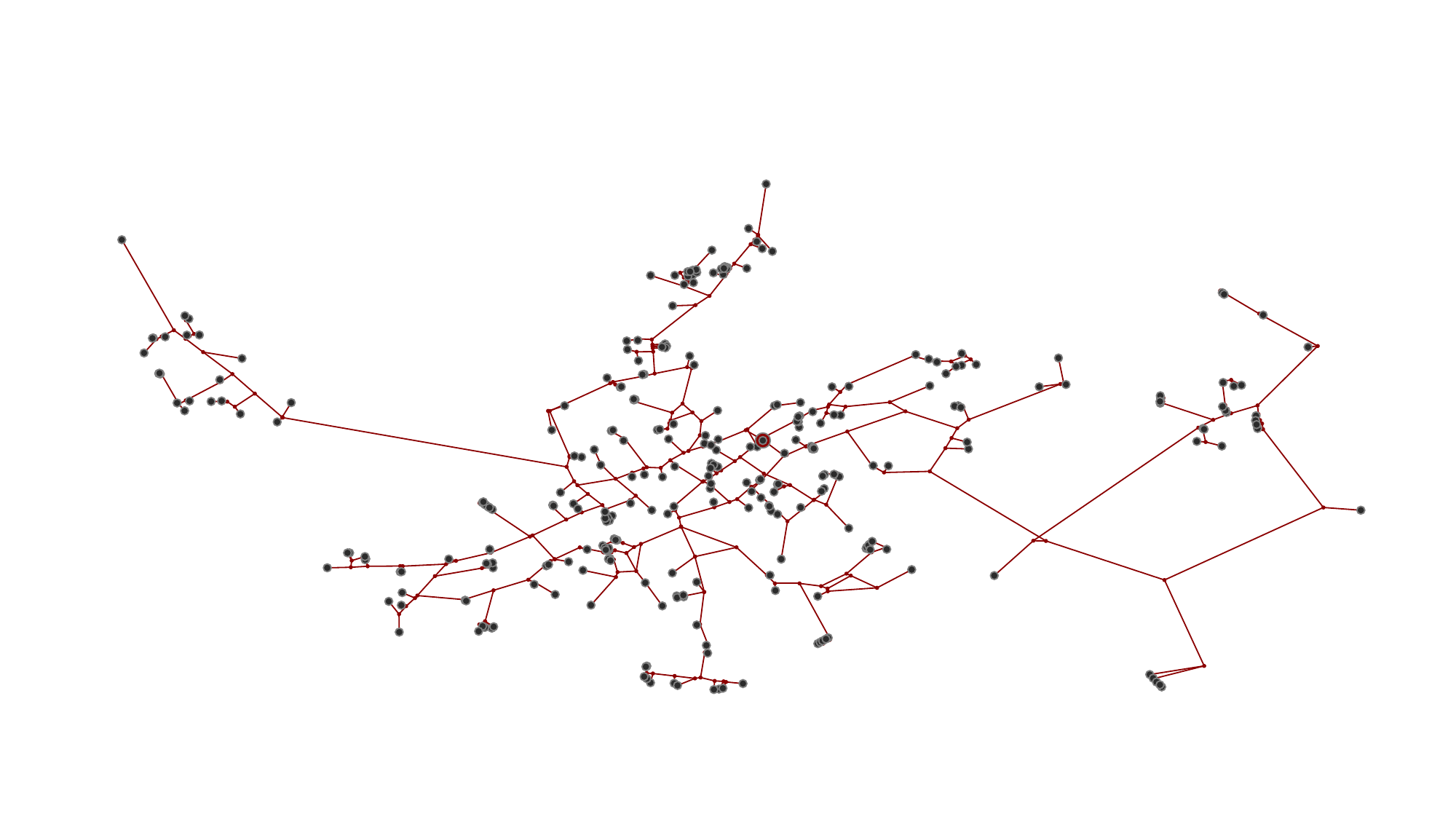}
    \caption{Clustering building nodes to aggregated consumer nodes; determining smaller number of clusters.}
    \label{fig_dh_model_step_9}
\end{figure}


%% file: DH_model_generation.bbl
\begin{thebibliography}{10}
\expandafter\ifx\csname url\endcsname\relax
  \def\url#1{\texttt{#1}}\fi
\expandafter\ifx\csname urlprefix\endcsname\relax\def\urlprefix{URL }\fi
\expandafter\ifx\csname href\endcsname\relax
  \def\href#1#2{#2} \def\path#1{#1}\fi

\bibitem{Li.2018}
H.~Li, N.~Nord, {Transition to the 4th generation district heating -
  possibilities, bottlenecks, and challenges}, {Energy Procedia} 149 (2018)
  483--498.
\newblock \href {https://doi.org/10.1016/j.egypro.2018.08.213}
  {\path{doi:10.1016/j.egypro.2018.08.213}}.

\bibitem{Lund.2018}
H.~Lund, P.~A. {\O}stergaard, M.~Chang, S.~Werner, S.~Svendsen, P.~Sorkn{\ae}s,
  J.~E. Thorsen, F.~Hvelplund, B.~O.~G. Mortensen, B.~V. Mathiesen, C.~Bojesen,
  N.~Duic, X.~Zhang, B.~M{\"o}ller, {The status of 4th generation district
  heating: Research and results}, {Energy} 164 (2018) 147--159.
\newblock \href {https://doi.org/10.1016/j.energy.2018.08.206}
  {\path{doi:10.1016/j.energy.2018.08.206}}.

\bibitem{Bloess.2018}
A.~Bloess, W.-P. Schill, A.~Zerrahn, {Power-to-heat for renewable energy
  integration: A review of technologies, modeling approaches, and flexibility
  potentials}, {Applied Energy} 212 (2018) 1611--1626.
\newblock \href {https://doi.org/10.1016/j.apenergy.2017.12.073}
  {\path{doi:10.1016/j.apenergy.2017.12.073}}.

\bibitem{Merlet.2023}
Y.~Merlet, R.~Baviere, N.~Vasset, {Optimal retrofit of district heating network
  to lower temperature levels}, {Energy} 282 (2023) 128386.
\newblock \href {https://doi.org/10.1016/j.energy.2023.128386}
  {\path{doi:10.1016/j.energy.2023.128386}}.

\bibitem{Brange.2017}
L.~Brange, P.~Lauenburg, K.~Sernhed, M.~Thern, {Bottlenecks in district heating
  networks and how to eliminate them -- A simulation and cost study}, {Energy}
  137 (2017) 607--616.
\newblock \href {https://doi.org/10.1016/j.energy.2017.04.097}
  {\path{doi:10.1016/j.energy.2017.04.097}}.

\bibitem{Guelpa.2023}
E.~Guelpa, M.~Capone, A.~Sciacovelli, N.~Vasset, R.~Baviere, V.~Verda,
  {Reduction of supply temperature in existing district heating: A review of
  strategies and implementations}, {Energy} 262 (2023) 125363.
\newblock \href {https://doi.org/10.1016/j.energy.2022.125363}
  {\path{doi:10.1016/j.energy.2022.125363}}.

\bibitem{Vannahme.2024}
A.~Vannahme, M.~Ehrenwirth, T.~Schrag, {Development and application of a
  guideline for assessing optimization potentials for district heating
  systems}, {Energy} 297 (2024) 131226.
\newblock \href {https://doi.org/10.1016/j.energy.2024.131226}
  {\path{doi:10.1016/j.energy.2024.131226}}.

\bibitem{Qin.2023}
Q.~Qin, L.~Gosselin, {Multiobjective optimization and analysis of
  low-temperature district heating systems coupled with distributed heat
  pumps}, {Applied Thermal Engineering} 230 (2023) 120818.
\newblock \href {https://doi.org/10.1016/j.applthermaleng.2023.120818}
  {\path{doi:10.1016/j.applthermaleng.2023.120818}}.

\bibitem{Hering.2022}
D.~Hering, M.~R. Faller, A.~Xhonneux, D.~M{\"u}ller, {Operational optimization
  of a 4th generation district heating network with mixed integer quadratically
  constrained programming}, {Energy} 250 (2022) 123766.
\newblock \href {https://doi.org/10.1016/j.energy.2022.123766}
  {\path{doi:10.1016/j.energy.2022.123766}}.

\bibitem{Pelda.2021}
J.~Pelda, S.~Holler, U.~Persson, {District heating atlas - Analysis of the
  German district heating sector}, {Energy} 233 (2021) 121018.
\newblock \href {https://doi.org/10.1016/j.energy.2021.121018}
  {\path{doi:10.1016/j.energy.2021.121018}}.

\bibitem{Muncan.2024}
V.~Mun{\'c}an, I.~Mujan, D.~Macura, A.~S. Andelkovi{\'c}, {The state of
  district heating and cooling in Europe - A literature-based assessment},
  {Energy} 304 (2024) 132191.
\newblock \href {https://doi.org/10.1016/j.energy.2024.132191}
  {\path{doi:10.1016/j.energy.2024.132191}}.

\bibitem{Triebs.2021}
M.~S. Triebs, E.~Papadis, H.~Cramer, G.~Tsatsaronis, {Landscape of district
  heating systems in Germany -- Status quo and categorization}, {Energy
  Conversion and Management: X} 9 (2021) 100068.
\newblock \href {https://doi.org/10.1016/j.ecmx.2020.100068}
  {\path{doi:10.1016/j.ecmx.2020.100068}}.

\bibitem{Manz.2024}
P.~Manz, A.~Billerbeck, A.~K{\"o}k, M.~Fallahnejad, T.~Fleiter, L.~Kranzl,
  S.~Braungardt, W.~Eichhammer, {Spatial analysis of renewable and excess heat
  potentials for climate-neutral district heating in Europe}, {Renewable
  Energy} 224 (2024) 120111.
\newblock \href {https://doi.org/10.1016/j.renene.2024.120111}
  {\path{doi:10.1016/j.renene.2024.120111}}.

\bibitem{IqonyGmbH.2024}
{Iqony GmbH},
  \href{https://fernwaerme.iqony.energy/de/fernwaerme-fuer-sie/versorgungsgebiete-check}{{Fernw{\"a}rme
  von Iqony}} (2024).
\newline\urlprefix\url{https://fernwaerme.iqony.energy/de/fernwaerme-fuer-sie/versorgungsgebiete-check}

\bibitem{OpenStreetMapcontributors.2024}
{OpenStreetMap contributors},
  \href{https://www.openstreetmap.org/}{{OpenStreetMap (OSM)}} (2024).
\newline\urlprefix\url{https://www.openstreetmap.org/}

\bibitem{Fuchs.2016}
M.~Fuchs, J.~Teichmann, M.~Lauster, P.~Remmen, R.~Streblow, D.~M{\"u}ller,
  {Workflow automation for combined modeling of buildings and district energy
  systems}, {Energy} 117 (2016) 478--484.
\newblock \href {https://doi.org/10.1016/j.energy.2016.04.023}
  {\path{doi:10.1016/j.energy.2016.04.023}}.

\bibitem{Nord.2021}
N.~Nord, M.~Shakerin, T.~Tereshchenko, V.~Verda, R.~Borchiellini, {Data
  informed physical models for district heating grids with distributed heat
  sources to understand thermal and hydraulic aspects}, {Energy} 222 (2021)
  119965.
\newblock \href {https://doi.org/10.1016/j.energy.2021.119965}
  {\path{doi:10.1016/j.energy.2021.119965}}.

\bibitem{Gross.2021}
M.~Gross, B.~Karbasi, T.~Reiners, L.~Altieri, H.-J. Wagner, V.~Bertsch,
  {Implementing prosumers into heating networks}, {Energy} 230 (2021) 120844.
\newblock \href {https://doi.org/10.1016/j.energy.2021.120844}
  {\path{doi:10.1016/j.energy.2021.120844}}.

\bibitem{Han.2023}
P.~Han, H.~Hua, H.~Wang, J.~Shang, {A graphic partition method based on nodes
  learning for energy pipelines network simulation}, {Energy} 282 (2023)
  128179.
\newblock \href {https://doi.org/10.1016/j.energy.2023.128179}
  {\path{doi:10.1016/j.energy.2023.128179}}.

\bibitem{Zhong.2020}
W.~Zhong, J.~Chen, Y.~Zhou, Z.~Li, Z.~Yu, X.~Lin, {Investigation of optimized
  network splitting of large-scale urban centralized heating system operation},
  {Energy Reports} 6 (2020) 467--477.
\newblock \href {https://doi.org/10.1016/j.egyr.2020.02.012}
  {\path{doi:10.1016/j.egyr.2020.02.012}}.

\bibitem{Schmidt.2021}
J.~Schmidt, P.~Stange, {Optimization of district heating network design},
  {Energy Reports} 7 (2021) 97--104.
\newblock \href {https://doi.org/10.1016/j.egyr.2021.09.034}
  {\path{doi:10.1016/j.egyr.2021.09.034}}.

\bibitem{Salenbien.2023}
R.~Salenbien, Y.~Wack, M.~Baelmans, M.~Blommaert, {Geographically informed
  automated non-linear topology optimization of district heating networks},
  {Energy} 283 (2023) 128898.
\newblock \href {https://doi.org/10.1016/j.energy.2023.128898}
  {\path{doi:10.1016/j.energy.2023.128898}}.

\bibitem{Li.2022}
X.~Li, A.~Walch, S.~Yilmaz, M.~Patel, J.~Chambers, {Optimal spatial resource
  allocation in networks: Application to district heating and cooling},
  {Computers {\&} Industrial Engineering} 171 (2022) 108448.
\newblock \href {https://doi.org/10.1016/j.cie.2022.108448}
  {\path{doi:10.1016/j.cie.2022.108448}}.

\bibitem{enercityAG.2024}
{enercity AG},
  \href{https://www.enercity.de/fernwaerme#karte}{{Fernw{\"a}rmesatzungsgebiet
  Hannover: Finden Sie heraus, ob Ihr Geb{\"a}ude innerhalb des
  Fernw{\"a}rmesatzungsgebietes liegt.}} (2024).
\newline\urlprefix\url{https://www.enercity.de/fernwaerme#karte}

\bibitem{wesernetzBremenGmbH.2024}
{wesernetz Bremen GmbH},
  \href{https://www.wesernetz.de/fuer-mein-zuhause/nachhaltiges-zuhause/fernwaerme-verfuegbarkeit}{{Fernw{\"a}rme-Verf{\"u}gbarkeit
  Bremen}} (2024).
\newline\urlprefix\url{https://www.wesernetz.de/fuer-mein-zuhause/nachhaltiges-zuhause/fernwaerme-verfuegbarkeit}

\bibitem{STAWAGStadtundStadteregionswerkeAachenAG.2011}
{STAWAG -- Stadt- und St{\"a}dteregionswerke Aachen AG} (2011).
\newblock
  \href{https://www.aachen-hat-energie.de/STAWAG/Plan_Fernwaerme.pdf}{[link]}.
\newline\urlprefix\url{https://www.aachen-hat-energie.de/STAWAG/Plan_Fernwaerme.pdf}

\bibitem{HausundGrundeigentumervereinHamburgRahlstedte.V..2023}
{Haus- und Grundeigent{\"u}merverein Hamburg-Rahlstedt e.V.},
  \href{https://www.hug-rahlstedt.de/wp-content/uploads/2023/03/Fernwaermesystemplan.pdf}{{Fernwaermesystemplan
  Hamburg}} (2023).
\newline\urlprefix\url{https://www.hug-rahlstedt.de/wp-content/uploads/2023/03/Fernwaermesystemplan.pdf}

\bibitem{FernwarmeUlmGmbH.2024}
{Fernw{\"a}rme Ulm GmbH},
  \href{https://www.fernwaerme-ulm.de/energie/versorgungsnetz/}{{Versorgungsnetz
  Ulm}} (2024).
\newline\urlprefix\url{https://www.fernwaerme-ulm.de/energie/versorgungsnetz/}

\bibitem{RheinEnergieAG.2024}
{RheinEnergie AG},
  \href{https://www.rheinenergie.com/de/privatkunden/waerme___wasser/waerme/fernwaerme_beziehen/fernwaerme_beziehen.html}{{Fernw{\"a}rme
  in K{\"o}ln beziehen: Das K{\"o}lner Fernw{\"a}rmenetz}} (2024).
\newline\urlprefix\url{https://www.rheinenergie.com/de/privatkunden/waerme___wasser/waerme/fernwaerme_beziehen/fernwaerme_beziehen.html}

\bibitem{StadtwerkeKarlsruheGmbH.2024}
{Stadtwerke Karlsruhe GmbH},
  \href{https://geoportal.karlsruhe.de/stadtplan/index.html?webmap=616597c654ea41d19f4002c5e8af1897}{{Fernw{\"a}rme
  aus Karlsruhe: Verf{\"u}gbarkeit {\&} Ausbaustatus}} (2024).
\newline\urlprefix\url{https://geoportal.karlsruhe.de/stadtplan/index.html?webmap=616597c654ea41d19f4002c5e8af1897}

\bibitem{Remmen.2018}
P.~Remmen, M.~Lauster, M.~Mans, M.~Fuchs, T.~Osterhage, D.~M{\"u}ller, {TEASER:
  an open tool for urban energy modelling of building stocks}, {Journal of
  Building Performance Simulation} 11~(1) (2018) 84--98.
\newblock \href {https://doi.org/10.1080/19401493.2017.1283539}
  {\path{doi:10.1080/19401493.2017.1283539}}.

\bibitem{Stock.2024}
J.~Stock, A.~Xhonneux, D.~M{\"u}ller, {Optimisation of district heating network
  separation for the utilisation of heat source potentials}, {Energy} 303
  (2024) 131872.
\newblock \href {https://doi.org/10.1016/j.energy.2024.131872}
  {\path{doi:10.1016/j.energy.2024.131872}}.

\bibitem{Stock.2024c}
J.~Stock, A.~Xhonneux, D.~M{\"u}ller, {Optimisation of District Heating Network
  Separation: An Extended Approach for Partial Transformation of Large-Scale
  Network Structures}, {In Review} (2024).

\bibitem{LandesamtfurNaturUmweltundVerbraucherschutzNRW.2024}
{Landesamt f{\"u}r Natur, Umwelt und Verbraucherschutz NRW},
  \href{https://www.energieatlas.nrw.de/}{{Energieatlas NRW}} (2024).
\newline\urlprefix\url{https://www.energieatlas.nrw.de/}

\bibitem{oemofDeveloperGroup.2016}
{oemof Developer Group}, \href{https://github.com/oemof/demandlib}{oemof
  demandlib} (2016).
\newblock \href {https://doi.org/10.5281/zenodo.438786}
  {\path{doi:10.5281/zenodo.438786}}.
\newline\urlprefix\url{https://github.com/oemof/demandlib}

\bibitem{InWISInstitutfurWohnungswesenImmobilienwirtschaftStadtundRegionalentwicklungGmbH.2021}
{InWIS - Institut f{\"u}r Wohnungswesen, Immobilienwirtschaft, Stadt- und
  Regionalentwicklung GmbH},
  \href{https://www.energieatlas.nrw.de/site/raumwarmebereitstellung}{{W{\"a}rmebereitstellung
  im Energieatlas NRW}} (2021).
\newline\urlprefix\url{https://www.energieatlas.nrw.de/site/raumwarmebereitstellung}

\bibitem{StatistischesBundesamt.2011}
{Statistisches Bundesamt}, \href{https://www.zensus2011.de/}{{Zensus 2011:
  Bev{\"o}lkerungs- und Wohnungsz{\"a}hlung}} (2011).
\newline\urlprefix\url{https://www.zensus2011.de/}

\bibitem{Nussbaumer.April2017}
T.~Nussbaumer, S.~Thalmann, A.~Jenni, J.~K{\"o}del, {Planungshandbuch
  Fernw{\"a}rme} (April 2017).

\bibitem{Guelpa.2019}
E.~Guelpa, V.~Verda, {Compact physical model for simulation of thermal
  networks}, {Energy} 175 (2019) 998--1008.
\newblock \href {https://doi.org/10.1016/j.energy.2019.03.064}
  {\path{doi:10.1016/j.energy.2019.03.064}}.

\bibitem{Falay.2020}
B.~Falay, G.~Schweiger, K.~O'Donovan, I.~Leusbrock, {Enabling large-scale
  dynamic simulations and reducing model complexity of district heating and
  cooling systems by aggregation}, {Energy} 209 (2020) 118410.
\newblock \href {https://doi.org/10.1016/j.energy.2020.118410}
  {\path{doi:10.1016/j.energy.2020.118410}}.

\bibitem{Kotzur.2021}
L.~Kotzur, L.~Nolting, M.~Hoffmann, T.~Gro{\ss}, A.~Smolenko, J.~Priesmann,
  H.~B{\"u}sing, R.~Beer, F.~Kullmann, B.~Singh, A.~Praktiknjo, D.~Stolten,
  M.~Robinius, {A modeler's guide to handle complexity in energy systems
  optimization}, {Advances in Applied Energy} 4 (2021) 100063.
\newblock \href {https://doi.org/10.1016/j.adapen.2021.100063}
  {\path{doi:10.1016/j.adapen.2021.100063}}.

\end{thebibliography}
